\documentclass[lettersize,journal]{IEEEtran}
\usepackage{hyperref}
\usepackage{amsmath,amsfonts}
\usepackage{algorithmic}
\usepackage{array}
\usepackage[caption=false,font=normalsize,labelfont=sf,textfont=sf]{subfig}
\usepackage{textcomp}
\usepackage{stfloats}
\usepackage{url}
\usepackage{verbatim}
\usepackage{graphicx}
\usepackage{cite}
\usepackage{siunitx}
\usepackage{orcidlink}
\def\BibTeX{{\rm B\kern-.05em{\sc i\kern-.025em b}\kern-.08em
    T\kern-.1667em\lower.7ex\hbox{E}\kern-.125emX}}
\usepackage{balance}
\begin{document}
\title{Multirate State Space Models for End-to-End Processing of

Pulse Density Modulated Speech Signals}
\author{Ludovic Boulanger\orcidlink{0009-0005-6560-1098} and Sean U. N. Wood\orcidlink{0000-0002-6821-1619}
\thanks{L. Boulanger and S. U. N. Wood are with the Department of Electrical and Computer Engineering at the University of Sherbrooke, Québec, Canada (ludovic.boulanger@usherbrooke.ca, sean.wood@usherbrooke.ca).}
\thanks{This work was enabled in part by support from the Quebec Research Fund (https://doi.org/10.69777/365045), Ministry of Economy, Innovation and Energy, Calcul Québec (calculquebec.ca), the Digital Research Alliance of Canada (alliancecan.ca), and the University of Sherbrooke.}
\thanks{The code will be made available at

https://github.com/NECOTIS/ssm-speech-processing.git}

\thanks{This work has been submitted to the IEEE for possible publication. Copyright may be transferred without notice, after which this version may no longer be accessible.}}


\maketitle

\begin{abstract}
Deep neural networks (DNNs) based on state-space models (SSMs) are increasingly applied to speech processing, but typically operate on pulse-code-modulated (PCM) audio. This constrains deployment on low-power, always-on edge devices, which commonly use single-bit pulse-density-modulated (PDM) micro-electromechanical (MEMS) microphones for their noise robustness, low cost, and variable sampling rates that enable low-power operation. In fact, converting PDM to PCM requires low-pass filtering and decimation, imposing costly overhead on resource-constrained hardware. While prior works have attempted to process PDM signals directly, they require long training times and generalize poorly across sampling rates. In this paper, we show that the SSM has two key properties that remediate these issues: its continuous-time parametrization allows it to produce a consistent representation of the input audio signal, regardless of the modulation strategy and sampling rate, and its long-term memory enables this representation to be aggressively downsampled without needing any anti-aliasing operations. We then propose a novel end-to-end PDM speech processing architecture that uses an SSM to encode the input audio signal into a modulation- and sampling-rate-invariant latent representation. We show that our proposed architecture achieves robust speech classification and enhancement gains at low-power sampling-rates (\SI{512}{\kHz}) and similar performance to state-of-the-art algorithms operating on PCM data when tested on standard PDM sampling-rates of \SI{2}{\MHz}. Moreover, we show that the SSM's output can be downsampled by more than 65,000 times, thus significantly reducing the number of processing timesteps in downstream layers.
\end{abstract}


\section{Introduction}
\IEEEPARstart{s}{tate-space} models (SSMs) have become increasingly popular in recent years for modeling continuous-time long-range dependencies in signals \cite{gu2022efficiently}. This property has been shown to be highly effective in speech processing applications which require long-term memory \cite{gu2022efficiently, pmlr-v162-goel22a}. In fact, speech processing models based on SSMs are frequently being used as an alternative to the popular Transformer's multi-head self-attention (MHSA) mechanism due to their computational efficiency \cite{zhang_mamba_2025, ding_keyword_2026}. However, these models typically process pulse-code-modulated (PCM) audio \cite{zhu_lightweight_2025, pei25_interspeech}. This limits their applicability to low-power, always-on edge devices devices that typically use pulse density modulated (PDM) micro-electromechanical (MEMS) microphones for their noise robustness, low cost, and variable sampling rates that enable very low-power operation \cite{black_analog--digital_1999, de_la_rosa_sigma-delta_2011, knowles_sph0641lm4h1, arnaud_yarga_end--end_2025}.

PDM microphones encode audio using a highly oversampled single-bit stream \cite{kite_understanding_2012}. Typically, speech processing algorithms interfacing with PDM require a PDM-to-PCM conversion module as a preprocessing step before processing the incoming sound \cite{chatterjee_clearbuds_2022, hagen_neural_2022}.  This conversion consists of a series of cascaded filters and decimation operations, which add undesired computational complexity to the system \cite{pavan_understanding_2017}. Moreover, as PDM microphones can vary their sampling rate to reduce energy costs \cite{knowles_sph0641lm4h1}, algorithms must be able to efficiently handle variable input sample rates. 

To date, there are two main approaches to circumvent the computational overhead imposed by PDM microphones. The first approach uses a convolutional neural network (CNN) to combine filtering and decimation operations into a single module. This approach has shown good performance when used as an input block in a keyword spotting architecture, while being significantly more energy-efficient than standard filtering techniques due to its lower computational complexity and ressource requirements \cite{vitolo_new_2023}.

The second approach involves processing the PDM data end-to-end. The PDM data is fed directly into a DNN, which can then learn optimal filters in the first layer for the given task. This strategy has proven to be effective in both PDM keyword spotting tasks and PDM-to-PCM speech enhancement tasks without the need for an explicit PDM-to-PCM conversion module \cite{yarga24_interspeech, arnaud_yarga_end--end_2025}.

These approaches suffer from two main disadvantages. First, they are typically trained on PDM data at very high sampling rates. This makes the training process very long and memory-intensive due to the long sequence lengths. Second, they typically train on only one sample rate, which prevents the network from generalizing to other PDM sampling rates. Therefore, the solutions cannot take full advantage of the PDM's multiple power modes without being trained on each individually.

Thus, there is a clear need for an efficient speech processing architecture that can bypass the computational overhead of long PDM sequences while remaining compatible with the various sampling rates produced by different PDM operating modes. To fill this gap, we propose a DNN that uses an SSM as an encoder layer to exploit the fact that PCM and PDM signals share the same spectral content below the PCM Nyquist frequency \cite{kite_understanding_2012}. We show that the continuous-time modeling of the SSM enables it to extract common features from both encodings, thereby producing a modulation-invariant representation of the input. Moreover, we show that this representation can be efficiently downsampled without anti-aliasing to a fixed-rate coefficient sequence, making the downstream layers agnostic to both the input modulation and sampling rate. Overall, this proposed approach leads to the following contributions:
\begin{enumerate}
  \item We propose a PDM speech processing architecture that can be trained on \SI{16}{\kHz} PCM data and naturally generalizes to PDM data, thereby avoiding the long training times and large memory requirements of PDM training.
  \item We test our proposed architecture on keyword spotting and speech enhancement tasks and show that the proposed architecture generalizes to low-power sampling-rates (\SI{512}{\kHz}) and offers similar performance than the state-of-the-art (SOTA) PCM-based algorithms when tested on standard PDM sampling-rates of \SI{2}{\MHz}, thereby eliminating the need to train separate models for each mode.
\end{enumerate}

This paper is organized as follows: we begin by presenting relevant theory on PDM encoding and SSMs in Section \ref{sec:background}. In Section \ref{sec:methods}, we then build on this theory and present our proposed PDM speech processing architecture. We present our experimental setup in Section \ref{sec:setup} and validate our method on a keyword spotting task and a more complex speech enhancement task in Section \ref{sec:experiments}, before concluding in Section \ref{sec:conclusion}.

\section{Background}\label{sec:background}

\begin{figure}[b!]
    \centering
    \includegraphics[width=\linewidth]{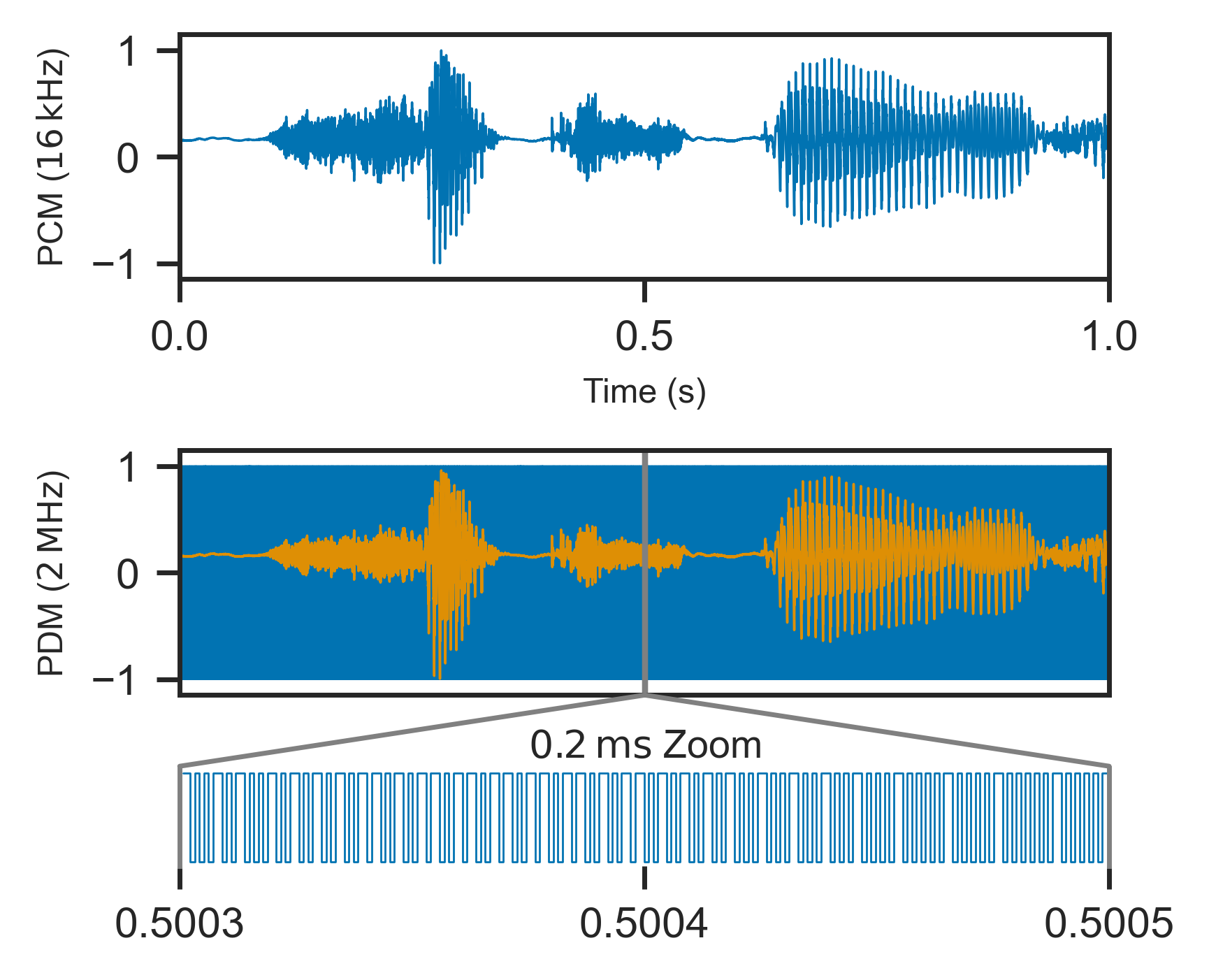}
    \caption{Comparison of a PCM encoded signal (top) and the same PDM encoded signal (middle and bottom). The orange overlay on the PDM data represents the low-pass-filtered PDM and demonstrates that the PCM baseband is preserved in the low-frequency range of the PDM. The low-pass was created using a 128-tap FIR filter with a cutoff frequency of \SI{7.5}{\kHz}. The bottom plot shows a \SI{200}{\us} zoom of the PDM encoded signal.}
    \label{fig:pdm-pcm}
\end{figure}

\subsection{Pulse Density Modulation}\label{sec:background-pdm}
PDM microphones can offer higher resolution for a lower energy cost than their PCM counterparts for low-bandwidth signals \cite{de_la_rosa_sigma-delta_2011}. This makes them ideal for speech processing, given that most of its energy is concentrated under \SI{8}{\kHz} \cite{french_factors_1947}. Moreover, PDM microphones often provide several operating modes by varying their sampling rate, such as a sleep mode (typically below \SI{350}{\kHz}), a low-power mode (typically between \SI{350}{\kHz} and \SI{1}{\MHz}), and a standard/performance mode (typically above \SI{1}{\MHz}), allowing them to operate at very low power when a high-resolution signal is not needed \cite{knowles_sph0641lm4h1, infineon_im70d122, puiaudio_dmm4326, st_mp23db01hp}.

These microphones use a single-bit Sigma-Delta ($\Sigma\Delta$) analog-to-digital converter (ADC) that samples the analog waveform at a much higher rate than the Nyquist rate (twice the highest encodable frequency in the input signal without introducing aliasing). The ratio between the sampling rate and the Nyquist rate is referred to as the oversampling ratio (OSR). The resulting PDM signal is a single-bit stream in which the temporal density of ones encodes the amplitude of the input signal \cite{kite_understanding_2012, pavan_understanding_2017}. This is in direct contrast to Nyquist ADCs that sample the incoming signal at the Nyquist rate and generate a multi-bit (e.g., 16-bit) representation of the input waveform at each sample time \cite{pavan_understanding_2017}. 

Figure \ref{fig:pdm-pcm} compares PCM and PDM encodings and highlights a key property of PDM signals: PCM data can be recovered by applying a lowpass filter and decimation to the PDM data. This is possible because both encodings share the same frequency content below the PCM Nyquist frequency \cite{kite_understanding_2012}, which is a consequence of the oversampling and noise shaping performed by the $\Sigma\Delta$ ADC \cite{de_la_rosa_sigma-delta_2011}.

Indeed, when Nyquist-rate ADCs quantize a continuous-time (CT) waveform, all the quantization noise lies in the signal band. By contrast, by oversampling the waveform, the $\Sigma\Delta$ ADC spreads the quantization noise across a much larger frequency range, resulting in less noise in the signal band. In addition, the $\Sigma\Delta$ ADC further reduces the quantization noise in the signal band by shaping it so that most of the noise is pushed above the PCM Nyquist frequency \cite{rosa_sigma-delta_2019}.

\begin{figure}[b!]
    \centering
    \includegraphics[width=\linewidth]{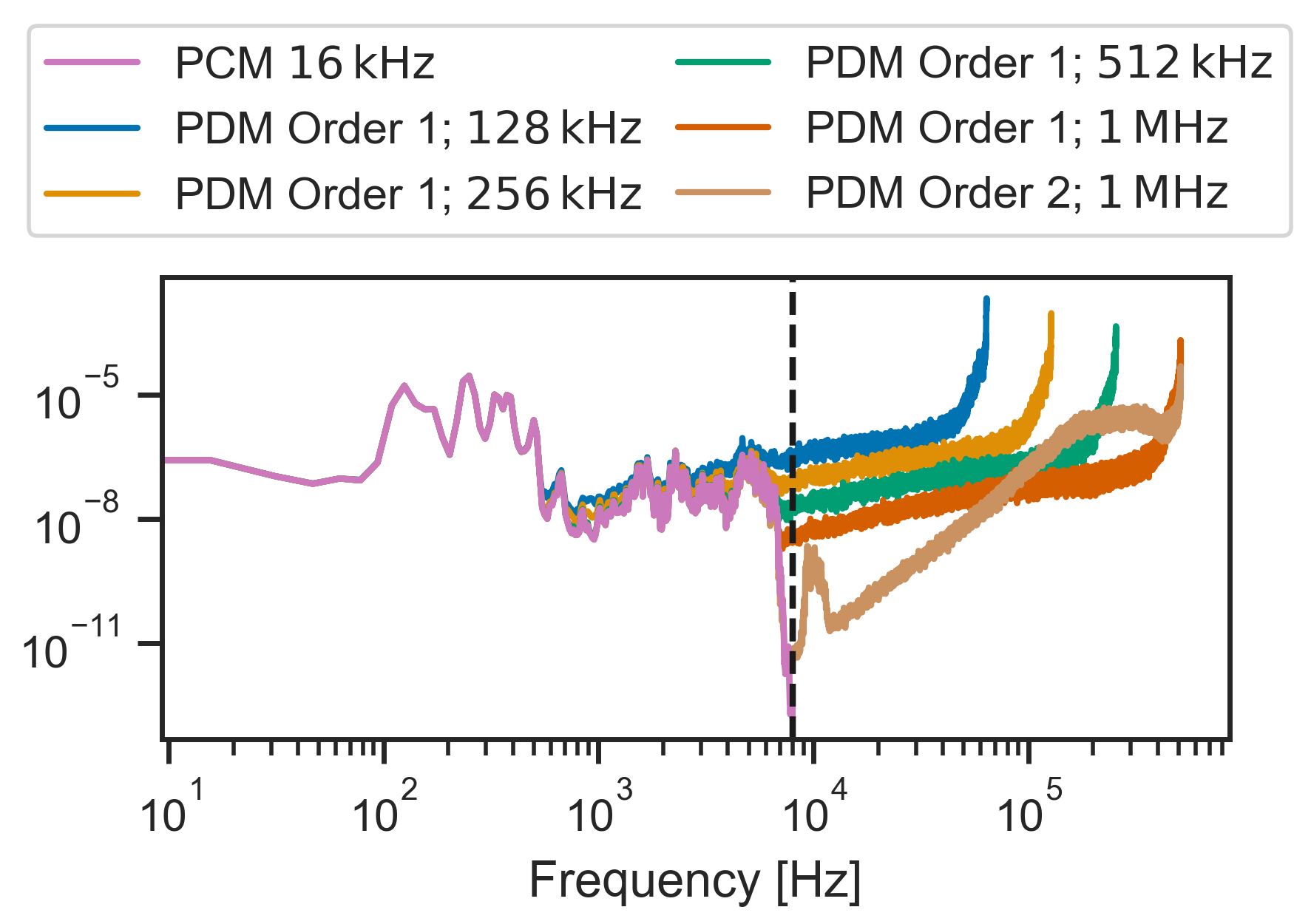}
    \caption{Comparison of the PDM and PCM power spectral densities (PSD) for a 1-second speech segment. The black dotted line represents the PCM Nyquist frequency of \SI{8}{\kHz}, i.e., the maximum frequency present in the signal. The pink line represents the PCM PSD.}
    \label{fig:pdm-spectra}
\end{figure}

However, in its most basic form, the $\Sigma\Delta$ ADC requires very high OSRs to achieve a given effective number of bits (ENOB), i.e., the resolution required by an ideal Nyquist ADC to match the dynamic range of an ideal $\Sigma\Delta$ ADC. In addition, the $\Sigma\Delta$ ADC can introduce structured quantization noise, known as idle tones, within the signal band \cite{de_la_rosa_sigma-delta_2011}. The performance of the $\Sigma\Delta$ ADC can be largely improved by cascading two $\Sigma\Delta$ modulators to create a second-order $\Sigma\Delta$ ADC \cite{rosa_sigma-delta_2019}.

The effects of oversampling, noise shaping, and modulator order are shown in Figure \ref{fig:pdm-spectra}, which compares the power spectral density (PSD) of various $\Sigma\Delta$ ADC configurations against the PCM PSD. We see that as the OSR increases, the $\Sigma\Delta$ PSD more closely follows the PCM PSD below the PCM Nyquist rate, indicating a cleaner signal band. Moreover, for an OSR of 64, the second-order $\Sigma\Delta$ PSD follows that of the PCM all the way up to the PCM Nyquist frequency of \SI{8}{\kHz}.

Overall, the efficiency of their underlying $\Sigma\Delta$ ADC, combined with their ability to sample at various rates, makes PDM microphones ideal for low-power and high-resolution speech processing in small, resource-constrained devices. In these cases, PDM microphones can offer higher resolution while consuming less power than their Nyquist counterparts due to their simpler electrical components \cite{de_la_rosa_sigma-delta_2011}. 

\subsection{State Space Models}\label{sec:background-ssms}

\begin{figure*}[!b]
 \center
  \includegraphics[width=\textwidth]{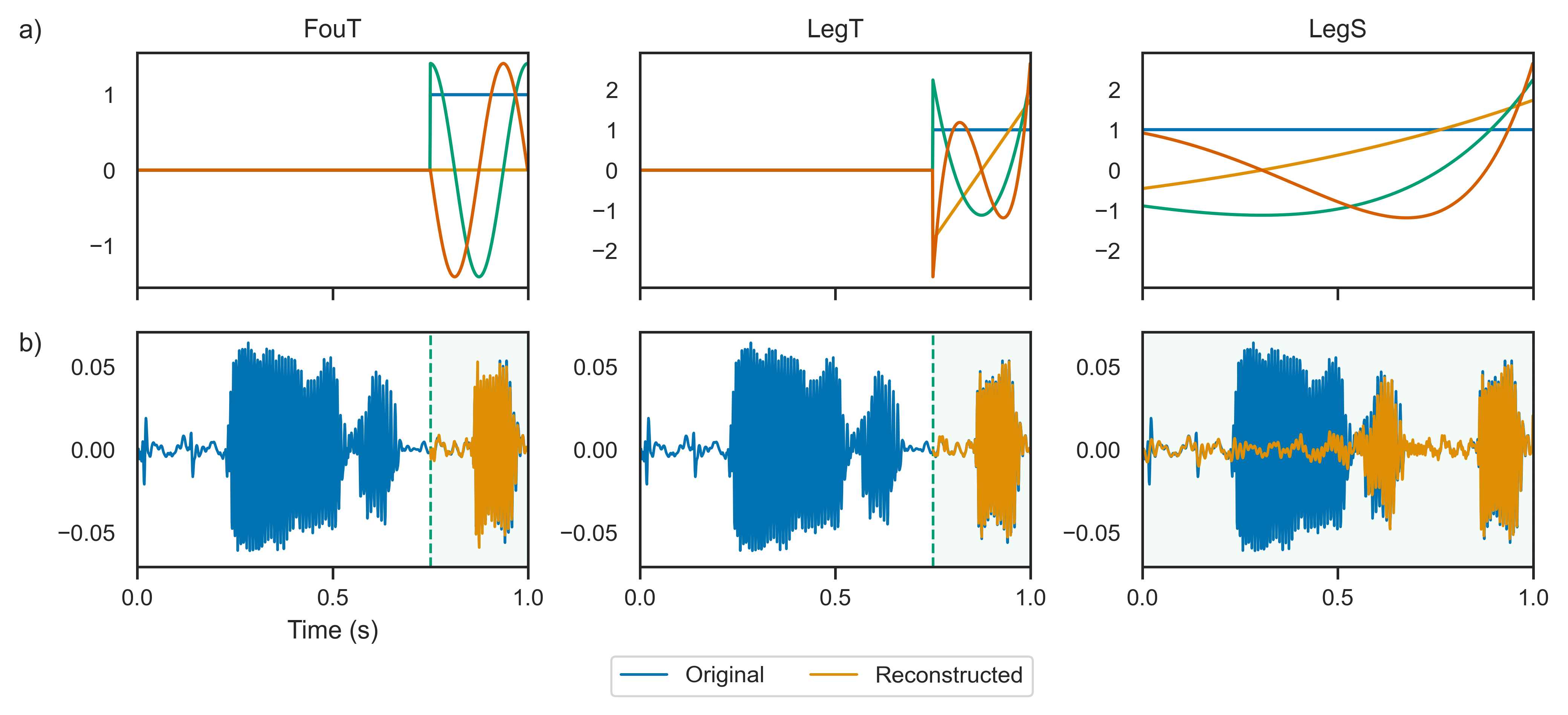}
  \caption{Demonstration of the basis functions and the memory windows of the FouT, LegT, and LegS SSMs. The SSMs all have a state dimension $H$ of 1024 and a discretization step $\Delta$ of 0.001 (\SI{16}{\kHz} sampling rate). The FouT and the LegT SSMs have a memory window size $\theta$ of \SI{250}{\ms} (4000 samples at \SI{16}{\kHz}). a) The first 4 basis functions used by FouT, LegT, and LegS. b) The reconstruction capabilities of the three initializations. The memory window length is denoted by the green overlay, and the original and reconstructed signals are represented by the blue and orange lines, respectively. The original signal is a \SI{16}{\kHz} speech signal lowpassed using a 1024-tap FIR filter with a cutoff frequency of \SI{100}{\Hz} for visualization purposes.}
  \label{fig:ssms}
\end{figure*}

SSMs have been applied to various fields such as engineering to design controllers for power systems \cite{prakash_design_2020}, economics to analyze the impact of oil prices \cite{nazif_catik_time-varying_2020}, and neuroscience to measure the progression of Parkinson's disease \cite{evers_measuring_2019}. Their application to deep learning (DL) began with the problem of designing the optimal memory unit capable of memorizing long-range dependencies without suffering from the vanishing gradient problem \cite{bengio_learning_1994}. This led to the appearance of the Legendre Memory Unit (LMU) \cite{NEURIPS2019_952285b9}, which was further generalized by the higher-order polynomial projection operators (HiPPO) framework \cite{gu_hippo_2020}.

In their most simple form, SSMs used as memory units are dynamical systems that memorize the past by mapping an input sequence $u(t)$ to a high-dimensional vector of coefficients $\mathbf{x}(t)$ that represents $\theta$ seconds of the input signal's history. This can be formally written as Equation \ref{eq:ssm}:
\begin{equation}
\begin{aligned}
\mathbf{\dot{x}}(t) &=  A\mathbf{x}(t) + Bu(t)
\end{aligned}
\label{eq:ssm}
\end{equation}
where $A\in\mathbb{R}^{H\times H}$, $B\in \mathbb{R}^{H\times 1}$ are the state and projection matrices respectively, and $H$ is the SSM's state dimension and determines the number of basis functions used to represent the history of the input. In this case, $\theta$ is a CT parameter that represents the SSM's memory duration in seconds. The coefficients $\mathbf{x}(t)$ enable the optimal reconstruction of $u(t)$ in the interval $[t-\theta, t]$ when projected onto a set of basis functions that span the duration of the memory.

SSMs are CT systems by nature. Thus, before being applied to a discrete-time (DT) signal, the $A$ and $B$ matrices need to be discretized by a step size $\Delta$ using techniques such as the bilinear method \cite{gu2022efficiently}. The resulting DT parameters $\bar{A}$ and $\bar{B}$ are given below:
\begin{equation}
\begin{aligned}
\bar{A} &= \left(I - \frac{\Delta}{2} A \right)^{-1} \left(I + \frac{\Delta}{2} A \right) \\
\bar{B} &= \left(I - \frac{\Delta}{2} A \right)^{-1} \Delta B \\
\end{aligned}
\label{eq:discrete-params}
\end{equation}
The DT SSM equation then becomes:
\begin{equation}
\begin{aligned}
\mathbf{x}[n] &=  \bar{A}\mathbf{x}[n-1] + \bar{B}u[n]
\end{aligned}
\label{eq:ssm-discrete}
\end{equation}
where $\bar{A}$ and $\bar{B}$ are the discretized versions of $A$ and $B$ respectively and $n$ corresponds to the DT sample index. 

The input's history can then be reconstructed using a linear combination of the chosen basis functions in which each element in the state vector $\mathbf{x}[n]$ acts as a weight for a particular basis:
\begin{equation}
\hat{\mathbf{u}}_n = \mathbf{P}\mathbf{x}[n]
\label{eq:reconstruct-ssm}
\end{equation}
where $\hat{\mathbf{u}}_n\in \mathbb{R}^{\bar{\theta}}$ is the reconstructed history, $\mathbf{P}\in \mathbb{R}^{\bar{\theta}\times H}$ are the basis functions and $\bar{\theta}$ is the DT version of the $\theta$ parameter using the step size $\Delta$, measured in samples.

A critical step for the success of SSM-based models is the initialization of the $A$ and $B$ matrices such that the coefficient vector $\mathbf{x}[n]$ can optimally reconstruct the past inputs when projected onto a particular set of basis functions \cite{gu_hippo_2020, NEURIPS2019_952285b9}. The three most popular initializations for $A$ and $B$ are known as LegS, LegT, and FouT which differ in their basis functions and their memorization capacities depicted in Figure \ref{fig:ssms} \cite{gu2023how}. The top row shows the first 4 basis functions for each SSM, and the bottom row shows the systems' memorization capacity. On the one hand, LegT and FouT have a fixed CT memory window of length $\theta$, measured in seconds. However, the two SSMs differ in their basis functions. In fact, FouT uses a scaled version of the truncated Fourier bases while LegT uses a scaled version of the Legendre polynomials, which are given in Equation \ref{eq:reconstruct-fout} and Equation \ref{eq:legendre} respectively \cite{gu2023how, lima_lecture_2025}.
\begin{equation}
P_h(x) = \begin{cases}
cos(2\pi h x)& \text{if h is 0},\\
\sqrt{2}cos(2\pi h x)& \text{if h is even},\\
\sqrt{2}sin(2\pi h x)& \text{otherwise}
\end{cases}
\label{eq:reconstruct-fout}
\end{equation}
\begin{equation}
P_h(x)=(-1)^n(2h+1)\frac{1}{2^h h!}\frac{d^h}{dx^h}(x^2-1)^h
\label{eq:legendre}
\end{equation}
In either case, the systems can nearly perfectly memorize the input signal within their memorization window but completely forget it outside that window. 

LegS attempts to encode the entire input history by projecting it onto Legendre polynomial coefficients, where the basis functions are evaluated on an exponentially warped domain. This results in memory accuracy that decays exponentially over time.

In this work, we focus on the LegT and FouT initializations because their sliding-window memory closely mirrors the analysis–processing–synthesis structure of short-time Fourier transform (STFT)-based speech enhancement. With these initializations, we can decompose the input into a time-coefficient representation that can be processed directly and reconstructed in the time domain through an inverse operation analogous to the inverse STFT. As shown in Section \ref{sec:se-results}, this structure allows models that perform well with STFT representations to transfer naturally to SSM-based representations without loss of performance.

\section{Methods}\label{sec:methods}
This work proposes an architecture that can be trained on PCM data and tested on PDM data across different OSRs with minimal performance loss. To this end, we use an SSM encoder layer to produce a representation that is invariant to modulation format and sampling rate, allowing the downstream network to operate only on a single, unified representation. We first present the SSM encoder layer in detail, then describe how we integrate it into the proposed architecture.

\subsection{SSM Encoder Layer}
\begin{figure}[!t]
 \center
  \includegraphics[width=\linewidth]{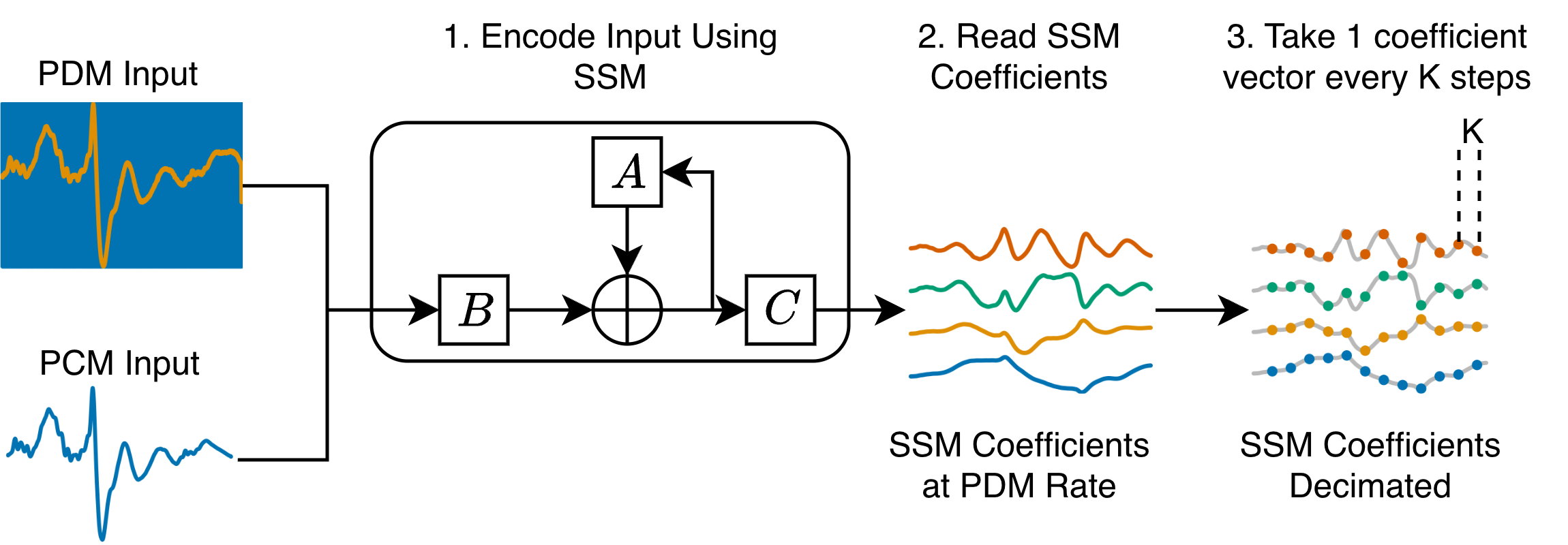}
  \caption{Our proposed SSM encoder layer. The SSM is responsible for creating a modulation- and sampling-rate-invariant representation of the input signal. To do so, it encodes the input data as a sequence of SSM coefficient vectors, then downsamples them from the variable-OSR PDM rate to a fixed rate by retaining one coefficient vector every $K=OSR$ time steps.}
  \label{fig:ssm-encoder}
\end{figure}

\begin{figure}[!b]
 \center
  \includegraphics[width=\linewidth]{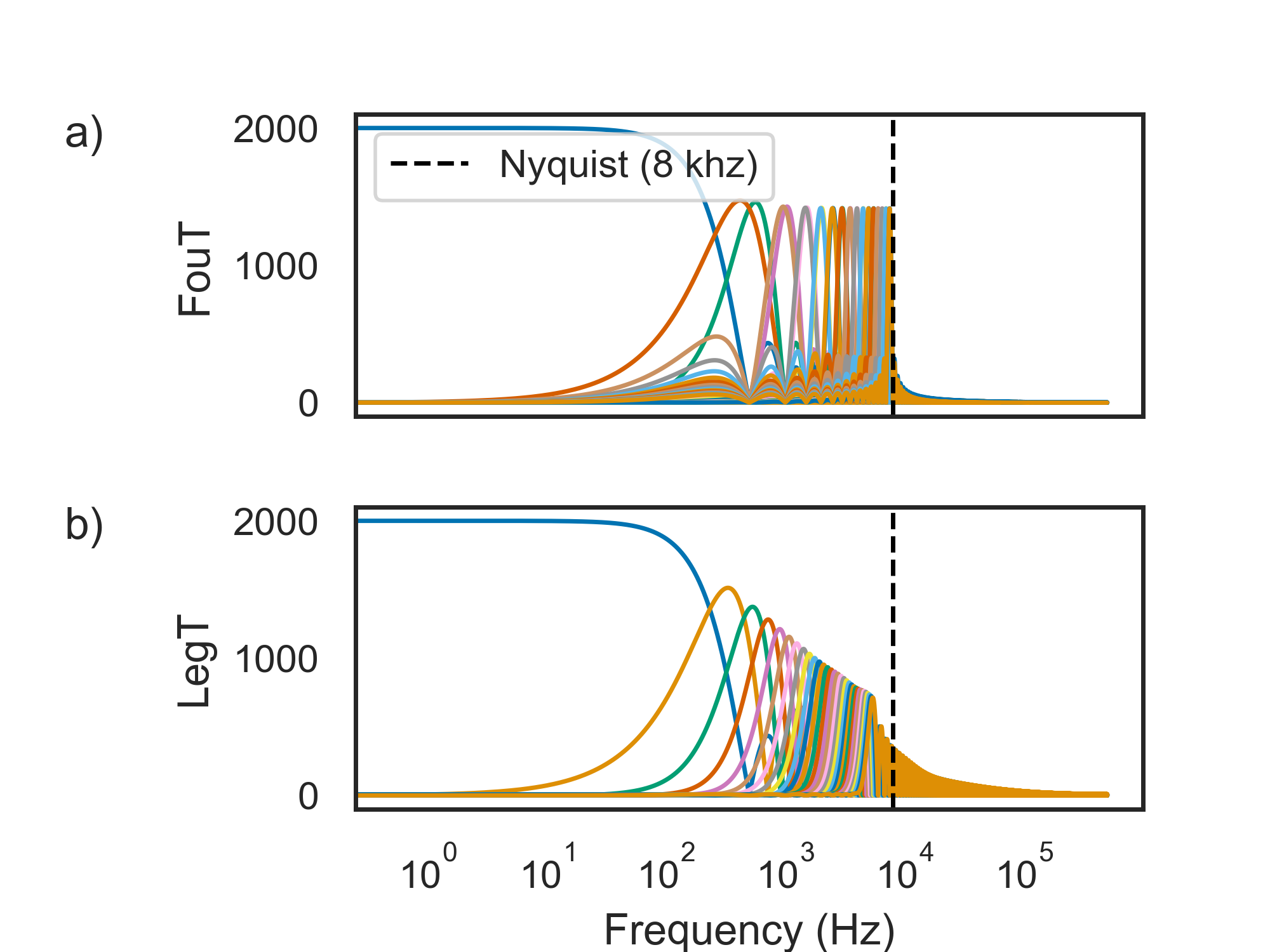}
  \caption{The frequency spectra of the FouT (a) and LegT (b) basis-functions for an SSM with $\theta=$ \SI{2}{\ms}, $H=32$. The black dotted line represents the \SI{16}{\kHz} PCM Nyquist frequency. For each SSM, the frequency spectrum was obtained by sampling each basis function at 1,000,000 uniformly distributed points over its domain and performing a 2,000,000-point DT Fourier transform.}
  \label{fig:ssm-passband}
\end{figure}

The SSM encoder layer, the core of our proposed architecture, is depicted in Figure \ref{fig:ssm-encoder}, which highlights its ability to convert both PCM and PDM sequences into a fixed-rate sequence for processing by downstream layers. To demonstrate how SSMs such as FouT and LegT create a modulation- and sampling-rate-invariant representation, we start by recalling from Section \ref{sec:background-pdm} that PDM and PCM modulations share the same frequency content below the PCM Nyquist frequency. As illustrated in Figure \ref{fig:ssm-passband}, FouT and LegT predominantly encode frequencies in this range. In fact, for both SSMs, most spectral energy lies below the \SI{8}{\kHz} PCM Nyquist frequency. The SSM response to PDM and PCM speech inputs is therefore expected to be very similar.

\begin{figure}[!t]
 \center
  \includegraphics[width=\linewidth]{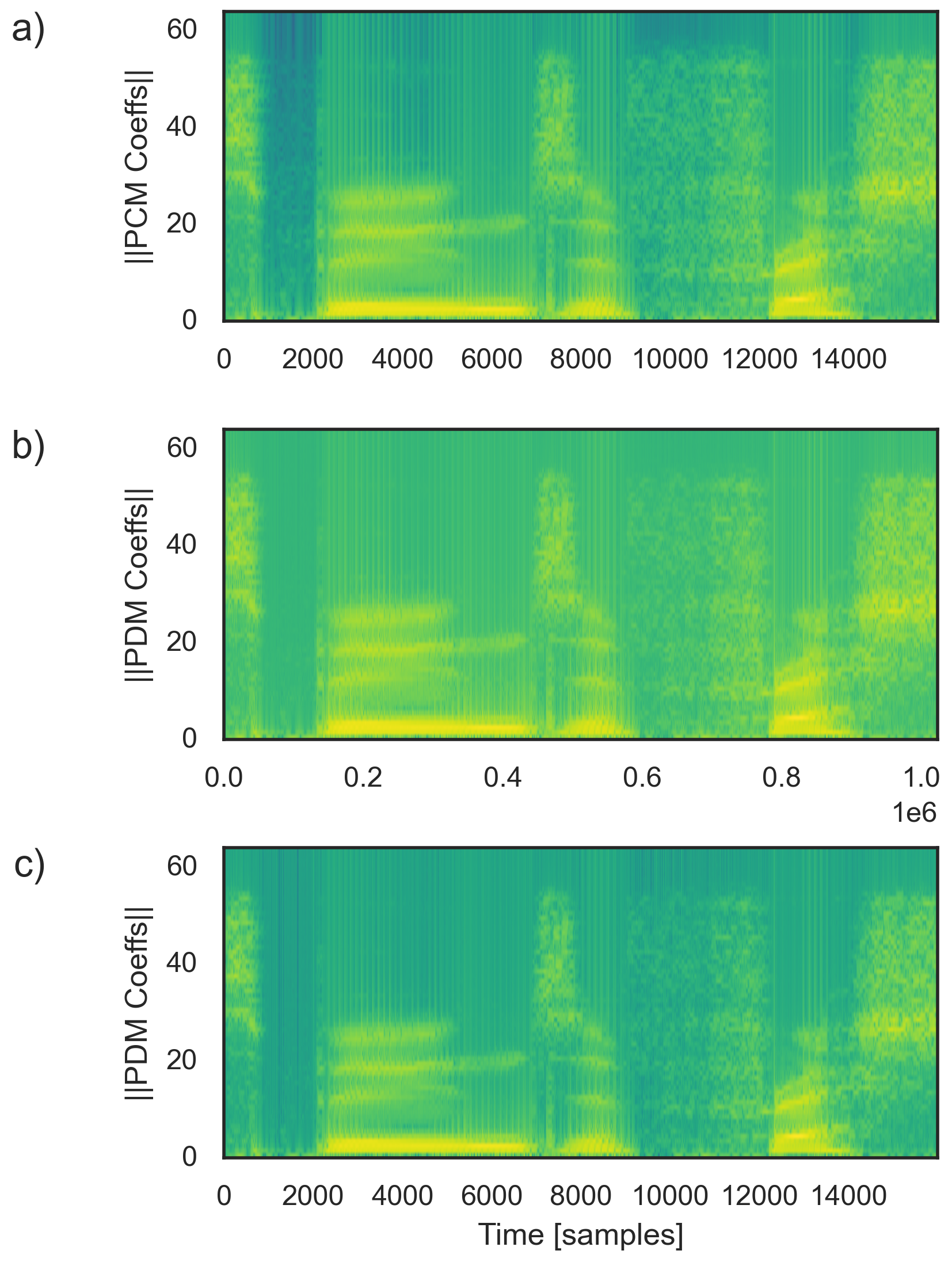}
  \caption{Comparison of the magnitude of the FouT coefficients SSM for a PCM and PDM input with $OSR=64$. The FouT SSM used a state dimension $H$ of 128 and a memory window size $\theta$ of \SI{8}{\ms}. The discretization step was set to $\Delta = 1/16,000$ and $\Delta=1/1,024,000$ for the PCM and PDM inputs respectively. a) The SSM coefficients sampled at the PCM rate, b) The SSM coefficients sampled at the PDM rate, and c) the SSM coefficients sampled at the PDM rate, then downsampled to the PCM rate by keeping only each 64th coefficient vector.}
  \label{fig:ssm-coeffs}
\end{figure}

Figures \ref{fig:ssm-coeffs}a) and b) illustrate this property for a FouT SSM encoding one second of speech represented either as \SI{16}{\kHz} PCM audio or as \SI{1.024}{\MHz} PDM audio respectively. The only difference between the two configurations is the discretization step, set to $\Delta = 1/16{,}000$ for PCM and $\Delta = 1/1{,}024{,}000$ for PDM. The resulting coefficient trajectories are nearly identical, up to the quantization noise introduced by the PDM modulation process, which is addressed in Section \ref{sec:pcm-training}.

However, as depicted by Figure \ref{fig:ssm-coeffs}b), the higher sampling rate of PDM also means that its SSM coefficients are produced at a proportionally higher rate than those of PCM. Subsequent layers trained only on PCM-derived coefficients may therefore fail to generalize to PDM-derived coefficients, since the temporal rate of the input representation is different. To resolve this mismatch, we exploit the CT memory window of LegT and FouT SSMs. Each state vector $\mathbf{x}[t] \in \mathbb{R}^{H}$ summarizes the input over the CT interval $[t-\theta, t]$, where $\theta$ is independent of the sampling rate. Consequently, the same memory window spans $\bar{\theta}_{\mathrm{PCM}}$ samples for PCM and $\bar{\theta}_{\mathrm{PDM}} = \mathrm{OSR}\times \bar{\theta}_{\mathrm{PCM}}$ samples for PDM. Moreover, adjacent SSM coefficient vectors therefore summarize highly overlapping CT windows, creating substantial temporal redundancy.

The PDM coefficient sequence can therefore be rate-matched to the PCM coefficient sequence by downsampling it with hop size $K_{\mathrm{PDM}} = \mathrm{OSR}$ without the need for any anti-aliasing operations. This is the third step of our SSM encoder depicted in Figure \ref{fig:ssm-encoder}. For integer $K_{\mathrm{PDM}}$, the downsampled PDM coefficients $\mathbf{x}[nK_{\mathrm{PDM}}]$ encode the same continuous-time history as the corresponding PCM coefficients $\mathbf{x}[n]$. This equivalence is visible in Figure~\ref{fig:ssm-coeffs}c), where the decimated PDM coefficients remains aligned with the non-decimated PCM coefficient spectrogram.

This coefficient downsampling technique can be extended by downsampling the PCM coefficient sequence below the Nyquist rate in the same way, further reducing the number of operations required by downstream modules without increasing the complexity of the overall method. We demonstrate the impact of this downsampling for keyword spotting and speech enhancement tasks in Sections \ref{sec:ks-downsampling} and \ref{sec:se-downsampling} respectively. If the PCM coefficients are downsampled by a factor $K_{PCM}$, the PDM coefficients must be downsampled by $K_{PDM} = K_{PCM} \times OSR$ to keep the encoder output rate consistent across modulations.

\subsection{Proposed PDM Speech Processing Architecture}

\begin{figure}[!t]
 \center
  \includegraphics[width=\linewidth]{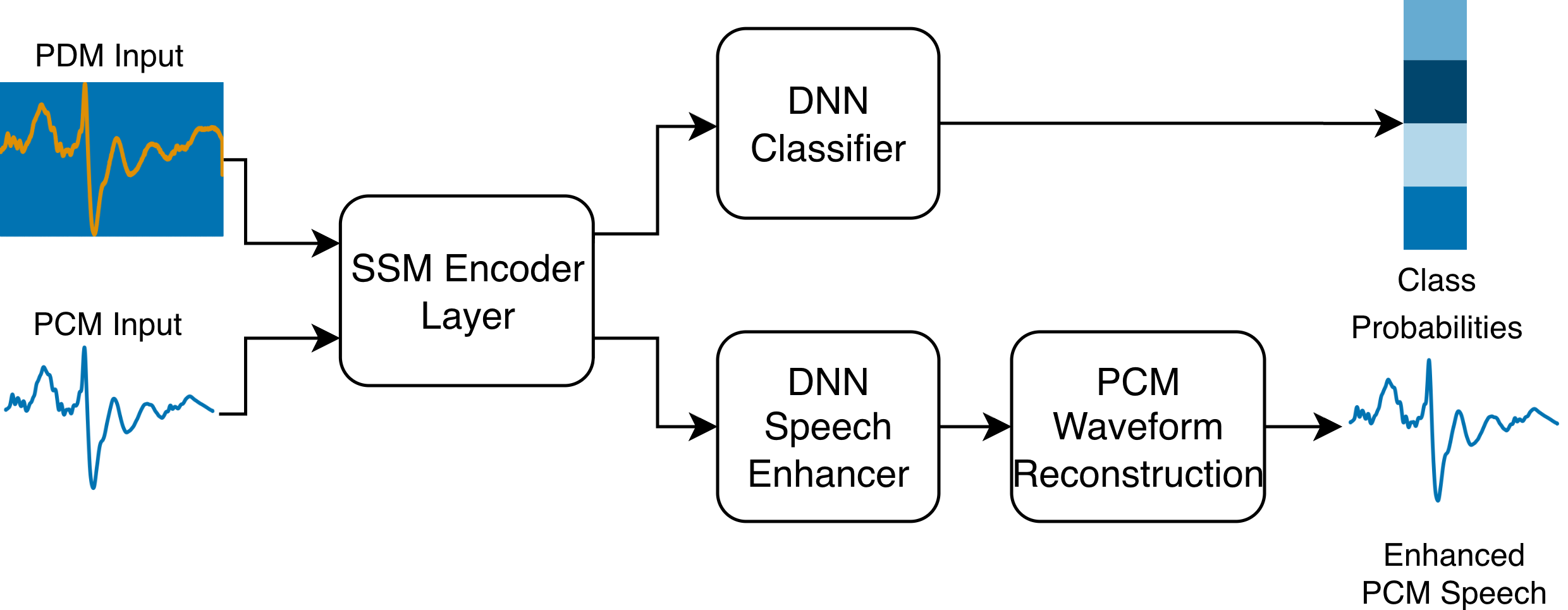}
  \caption{Our proposed PDM processing architecture. The SSM encoder layer is responsible for creating a modulation- and sampling-rate-invariant representation of the input signal. This representation can then be sent to a downstream DNN, which can be either a classifier for the keyword spotting task or a speech enhancement DNN. In the speech enhancement task, the predicted coefficients are used to reconstruct a clean time-domain PCM waveform.}
  \label{fig:architecture}
\end{figure}

The previously described SSM encoder layer enables an efficient PDM speech-processing architecture for both keyword spotting and speech enhancement. The proposed architecture is shown in Figure \ref{fig:architecture}.

The SSM encoder first maps the input speech signal to a sequence of SSM coefficients. For keyword spotting, this sequence of coefficients is passed to a classifier that predicts keyword class probabilities. For speech enhancement, the coefficients are instead passed to a speech-enhancement DNN, which predicts clean-speech SSM coefficients. The enhanced PCM waveform is then reconstructed using the procedure described in Section \ref{sec:background-ssms}: each predicted coefficient vector $\hat{\mathbf{x}}[n]$ is projected onto the SSM basis functions sampled at the PCM rate to produce a time-domain frame, and the resulting frames are combined by overlap-add (OLA) to reconstruct the full waveform.

\section{Experimental Setup}\label{sec:setup}
The proposed architecture represents an input signal as a sequence of modulation- and sampling-rate-invariant SSM coefficients. This enables existing PCM-specific keyword-spotting and speech enhancement DNNs to be extended to PDM data without substantially increasing training time or memory requirements. This section details the training setups used for both tasks, in which the DNNs are trained on PCM data and evaluated on their ability to generalize to PDM at various sample rates.

\subsection{PCM Training}\label{sec:pcm-training}
All training was done on \SI{16}{\kHz} PCM data. As demonstrated in Section \ref{sec:methods}, however, PDM encoding can add quantization noise in the signal band. The amount of added noise depends on the PDM's OSR and the $\Sigma\Delta$ modulator's order. Since we want our architecture to work for various PDM OSRs, we augmented the PCM training set with high-pass filtered white noise. The transfer function used for the filter is given in Equation \ref{eq:hp-filter}:
\begin{equation}\label{eq:hp-filter}
H(z) = G(1-\beta z^{-1})
\end{equation}
where $G$ is the gain of the filter and $\beta$ is a constant controlling how much noise leaks into the low frequencies mimicking the quantization noise introduced at low OSRs. During training, the gain $G$ is chosen from a log-uniform distribution ranging from $[10^{-3}, 10^{-1}]$. This allowed our architecture to see a range of noise levels during training and thus to better generalize across the various PDM OSRs. The $\beta$ parameter was fixed to 0.93. This value was empirically chosen so that the augmented noisy PCM data's power spectral density more closely matched that of the PDM data.

Specifically for the speech enhancement experiments, we added another augmentation technique that allows the network to learn to increase the effective bit rate between its input and output. This strategy is motivated by the fact that PDM signals sampled at \SI{128}{\kHz} have an effective bit-rate of \SI{128} kbps. On the other hand, \SI{16}{\kHz} PCM signals have an effective bit-rate of \SI{256} kbps (assuming 16-bit PCM). Thus, the target clean waveform has a higher resolution than the input PDM waveform when the OSR is very low. Hence, we added a quantization data-augmentation technique to the training process. This technique consisted of randomly selecting a quantization level between 8 and 16, corresponding to \SI{128} kbps and \SI{256} kbps, respectively, and quantizing the noisy waveform to that level. The target was kept untouched.

\subsection{PDM Evaluation}
To evaluate our models on PDM data, we first interpolate the PCM data using a cubic spline interpolation with a specific OSR factor and use a single-bit second-order $\Sigma\Delta$ modulator to convert the interpolated PCM stream to a PDM stream. We tested on OSR ranging from 8-128, representing sampling rates from \SI{128}{\kHz} to \SI{2}{\MHz}.

\subsection{Keyword Spotting Training Setup}
\subsubsection{Dataset}
The dataset used for this task is the Google Speech Commands Dataset \cite{warden_speech_2018}. This dataset consists of 105,829 clean English utterances from 2,618 speakers, split into training, validation, and testing subsets of 8,843, 9,981, and 11,005 utterances, respectively. The input audio is in \SI{16}{\kHz} PCM format, and the target is one of 35 keywords.

\subsubsection{Downstream DNN}
For the keyword spotting task, we integrated the SSM stack proposed by \cite{NEURIPS2022_e9a32fad} as the downstream DNN in our proposed architecture. This SSM stack consists of a series of LegS-initialized SSM blocks connected with dense layers. The number of SSM blocks used in this task is 6; each block uses a uni-directional LegS-initialized SSM as a memory cell with a hidden dimension of 64 and an input and output feature dimension of 64. All the parameters from Equation \ref{eq:ssm} were fine-tuned with gradient descent, and each channel had a unique learnable $\Delta$ parameter. Preliminary experiments showed that using a learnable $\Delta$ proved essential for each channel to model different timescales and significantly improved classification accuracy.

Finally, the output of the SSM blocks is averaged in time and sent to a dense layer that projects it into a 35-dimensional vector of logits for classification. The classified class was taken as the $\mathrm{argmax}$ of the output logits.

\subsubsection{Training Procedure}
Similar to \cite{NEURIPS2022_e9a32fad}, all models were trained for a total of 50 epochs using the AdamW \cite{loshchilov2018decoupled} optimizer with a learning rate of $10^{-2}$ and weight decay of 0.05, with the exception of the parameters in Equation \ref{eq:ssm}, which had a learning rate of $10^{-3}$ for stability. A cosine annealing learning rate scheduler was used to reduce the learning rate after every training epoch. Finally, the loss function used in this task is the cross-entropy loss between the predicted probabilities and the target one-hot vector.

\subsection{Speech Enhancement Training Setup}
\subsubsection{Dataset}
The dataset used for this task is the VoiceBank+Demand dataset \cite{valentini_dataset_2017, valentini2017}. This dataset has been used significantly in the literature and therefore serves as a good baseline for our validation \cite{yan_lisennet_2025}. The training set consists of 11,572 utterances from 28 speakers split equally into male and female speakers mixed with 8 noise types from the DEMAND dataset \cite{thiemann_diverse_2013} and 2 artificially created noises at signal-to-noise (SNR) ratios ranging from \SI{0}{\dB} to \SI{15}{\dB}. We also downsampled the original training audio from \SI{48}{\kHz} to \SI{16}{\kHz}. The testing set consists of 872 utterances from 2 new speakers (1 male and 1 female) mixed with 5 new noise types from the DEMAND dataset at SNRs ranging from \SI{2.5}{\dB} to \SI{17.5}{\dB}.

\subsubsection{Downstream DNN}
For the speech enhancement task, we integrated the LiSenNet DNN proposed in \cite{yan_lisennet_2025} as the downstream DNN within our proposed architecture. This model has shown remarkable speech enhancement capabilities while using under 40k parameters. We removed the Griffin-Lim algorithm from the architecture due to its iterative nature, which increases the overall algorithmic latency. We also removed the skip connections between the encoder and the decoder to reduce the total number of parameters in the network. Finally, the model's output consists of the SSM coefficients that enable reconstruction of the clean PCM waveform when projected onto their respective basis functions as described in Section \ref{sec:background-ssms}.

\subsubsection{Training Procedure}
The training procedure and training hyperparameters were kept the same as in \cite{yan_lisennet_2025}. Briefly, the training procedure used an adverserial training technique based on MetricGAN \cite{fu2019metricGAN}. On the one hand, the discriminator model was used to predict the perceptual evaluation of speech quality (PESQ) \cite{rix_perceptual_2001}, scaled between 0 and 1, from the log-compressed STFT magnitudes of the estimated and target speech. On the other hand, the generator was tasked to predict the clean waveform from the noisy waveform. Its loss function consisted of a weighted sum of the mean-squared error (MSE) between the log-compressed STFT spectrum of the estimated speech and the log-compressed STFT spectrum of the target speech, an MSE between the log-compressed STFT magnitude of the estimated speech and the log-compressed STFT magnitude of the target speech, and an MSE between the scaled PESQ score of the predicted speech and the PESQ score predicted by the discriminator. The weights for each term were set to 0.1, 0.9, and 0.05, respectively. Both the generator and discriminator models were trained with the AdamW optimizer \cite{loshchilov2018decoupled} using a learning rate of $10^{-3}$ and no weight decay.

In contrast to the original LiSenNet architecture, which predicted the clean STFT magnitudes and phases, our version predicts the clean SSM coefficients. Therefore, to facilitate integration with the original LiSenNet training setup, a 512-point STFT with a hop size of 256 was applied to the PCM waveform synthesized from the predicted SSM coefficients and the target \SI{16}{\kHz} PCM speech. This strategy allowed us to use the adversarial training from the original paper \cite{yan_lisennet_2025} without modification.

\section{Experimental Results}\label{sec:experiments}
We now present three experiments that both highlight the effectiveness of our proposed architecture and provide intuition for the choice of SSM hyperparameter values for a given task. The hyperparameters that will be tested are the memory window length $\theta$ (the length of history encoded by each SSM coefficient vector $\mathbf{x}[n]$), the decimation factor $K_{PDM}$ (the hop size between adjacent selected coefficient vectors) and the SSM state dimension $H$ (the number of basis functions used to encode the SSM's input). 

All experiments will be conducted on the keyword spotting task and the PDM-to-PCM speech enhancement task. The first experiment consists of fixing $K_{PDM}$ and varying $\theta$ and $H$. The second experiment consists of fixing $H$ and varying $K_{PDM}$ and $\theta$. These two experiments were chosen to test the effect of each SSM hyperparameter on keyword spotting and speech enhancement performance, and were conducted using \SI{2}{\MHz} PDM data, as this sampling rate is representative of a standard/high resolution PDM microphone mode. Finally, the third experiment is to fix all SSM hyperparameters, train the configuration on \SI{16}{\kHz} PCM waveforms, and test it across all PDM sample rates from \SI{128}{\kHz} to \SI{2}{\MHz}.

\subsection{Keyword Spotting Experiments}
\subsubsection{Effect of Varying the Memory Window Length $\theta$ and State Dimension $H$ on Classification Accuracy}
\begin{figure}[t!]
    \centering
    \includegraphics[width=\linewidth]{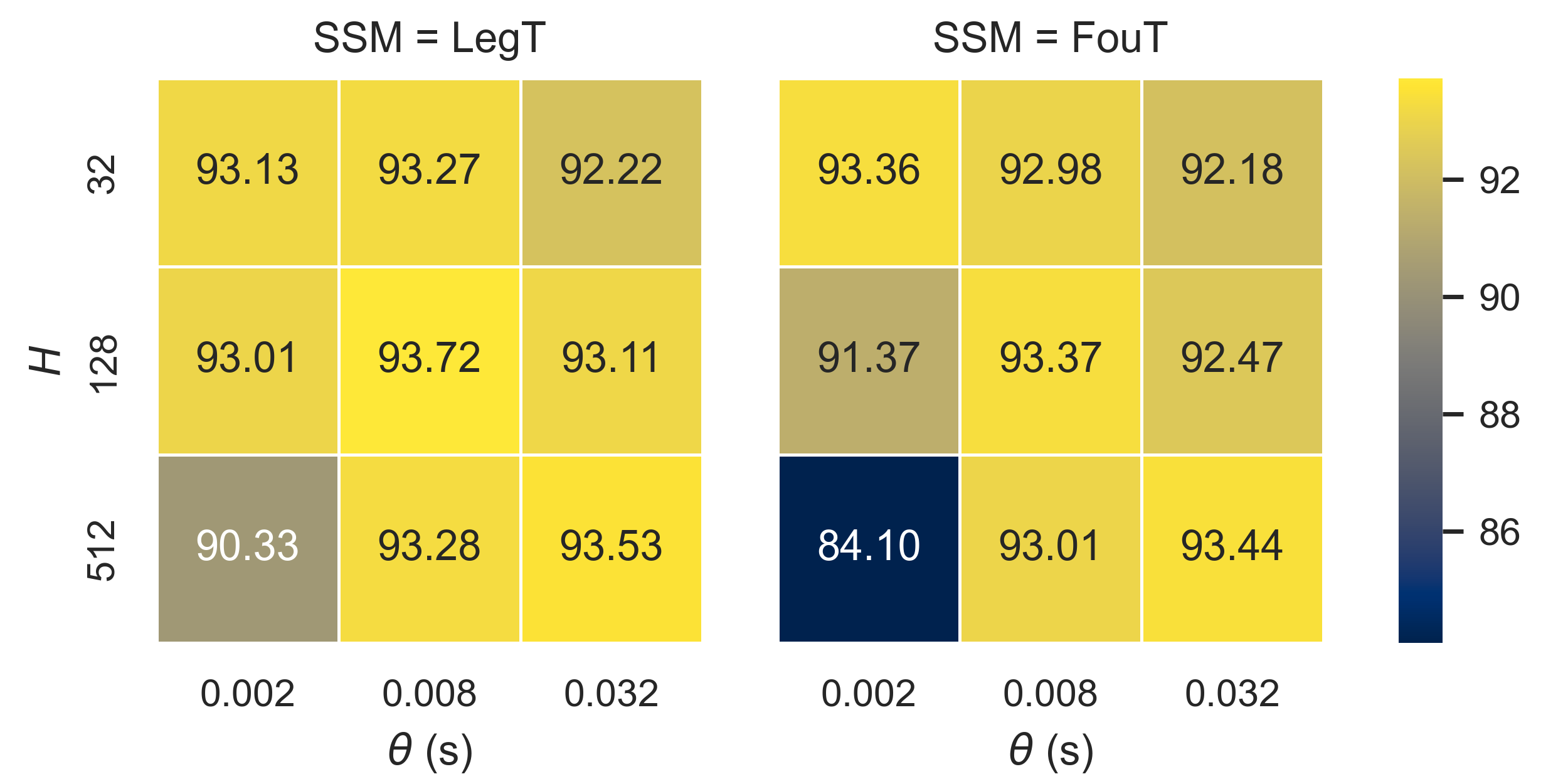}
    \caption{Effect of $H$ and $\theta$ on the classification results on the Google Speech Commands test set encoded using a single-bit second-order $\Sigma\Delta$ modulator with an OSR of 128 (\SI{2}{\MHz}). For both SSM families, the decimation factor $K_{PDM}$ was fixed at a value of 128, the window size $\theta$ was varied from \SI{2}-\SI{32}{\ms}, and the state dimension $H$ was varied from 32 to 512.}
    \label{fig:pdm-ks-results-window}
\end{figure}
The results for this experiment are shown in Figure \ref{fig:pdm-ks-results-window}. Out of all tested combinations, the LegT SSM with $H=128$ and $\theta=$ \SI{8}{\ms} offers the highest classification accuracy of 93.72\%. This result is higher than the best FouT combination, which offers 93.44\% accuracy with $H=512$ and $\theta=$ \SI{32}{\ms}. This implies that, for the classification task, the LegT SSM offers better performance using only a quarter of the bases than the FouT SSM.

Moreover, Figure \ref{fig:pdm-ks-results-window} shows that the difference in classification accuracy between $H=128$ and $H=32$ is very small for the LegT SSM. Since the SSM $A$ matrix scales quadratically with the state dimension $H$, using $H=32$ is significantly less computationally demanding than using $H=128$. This is promising for power- and resource-constrained devices, as they could use a much lighter 32-dimensional SSM without sacrificing keyword spotting accuracy.

We note that these results are slightly lower than the reported results in \cite{NEURIPS2022_e9a32fad} (around 96\%). This can be explained by the fact that we used a lower channel count and no bi-directional layers. In fact, even our largest SSM with $H=512$ results in a total parameter count of 112,000 parameters, which is significantly lower than the 307,000 reported by \cite{NEURIPS2022_e9a32fad}.

\subsubsection{Effect of the Decimation Factor $K_{PDM}$ and the Memory Window Length $\bar{\theta}_{PDM}$ on the Classification Accuracy}\label{sec:ks-downsampling}
\begin{figure}[b!]
    \centering
    \includegraphics[width=\linewidth]{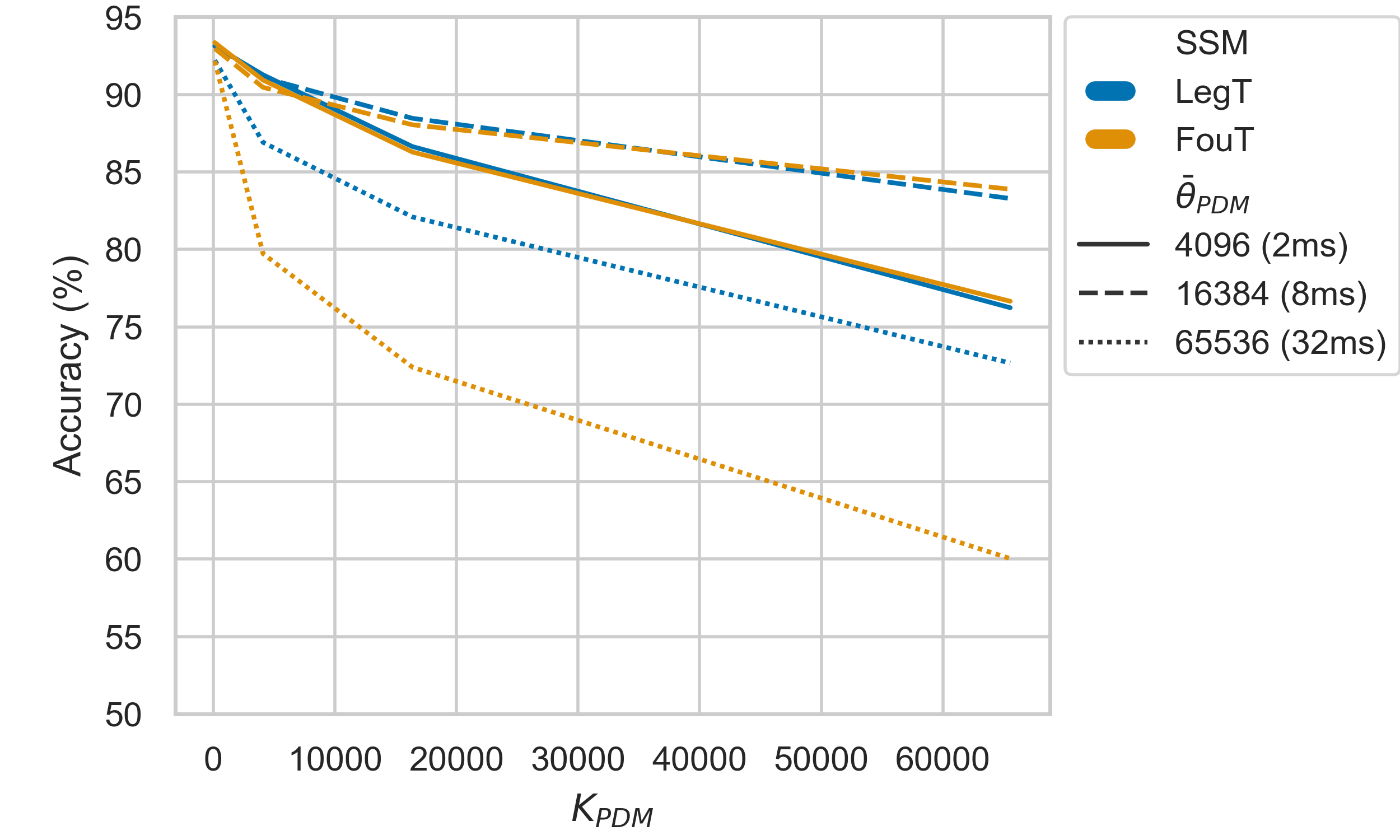}
    \caption{Effect of $K_{PDM}$ and $\bar{\theta}_{PDM}$ on the classification results on the Google Speech Commands test set encoded using a single-bit second-order $\Sigma\Delta$ modulator with an OSR of 128 (\SI{2}{\MHz}). For both SSM families, the decimation factor was varied from 128-65536, and the window size $\bar{\theta}_{PDM}$ was varied from 4,096-65,536 (\SI{2}-\SI{32}{\ms}). The state dimension $H$ was fixed at 128.}
    \label{fig:pdm-ks-results-decimation}
\end{figure}
Figure \ref{fig:pdm-ks-results-decimation} compares the LegT and FouT SSMs as a function of the PDM memory window size in samples $\bar{\theta}_{PDM}$ and decimation factor $K_{PDM}$. In this experiment, we fix the state dimension to $H=32$, which corresponds to the lowest-complexity configuration. For $\bar{\theta}_{PDM}=16,384$ (\SI{8}{\ms}), the proposed architecture maintains an accuracy of 88\% for decimation factors up to $K_{PDM}=\bar{\theta}_{PDM}$, and 83\% for decimation factors up to $K_{PDM}=4\times\bar{\theta}_{PDM}$. Overall, the accuracy drops by less than 10\% across all tested decimation factors.

These results show that the proposed architecture not only avoids the computational cost associated with the higher sampling rate of PDM signals compared to PCM signals but also enables further decimation of the SSM coefficients, substantially reducing the number of operations performed by the neural network without the need for any anti-aliasing operations. In fact, for a \SI{2}{\MHz} PDM sampling rate, our SSM encoder reduces this rate by a factor of 65,536, resulting in downstream layers processing data at only \SI{31.25}{\Hz}.

\begin{figure}[t!]
    \centering
    \includegraphics[width=\linewidth]{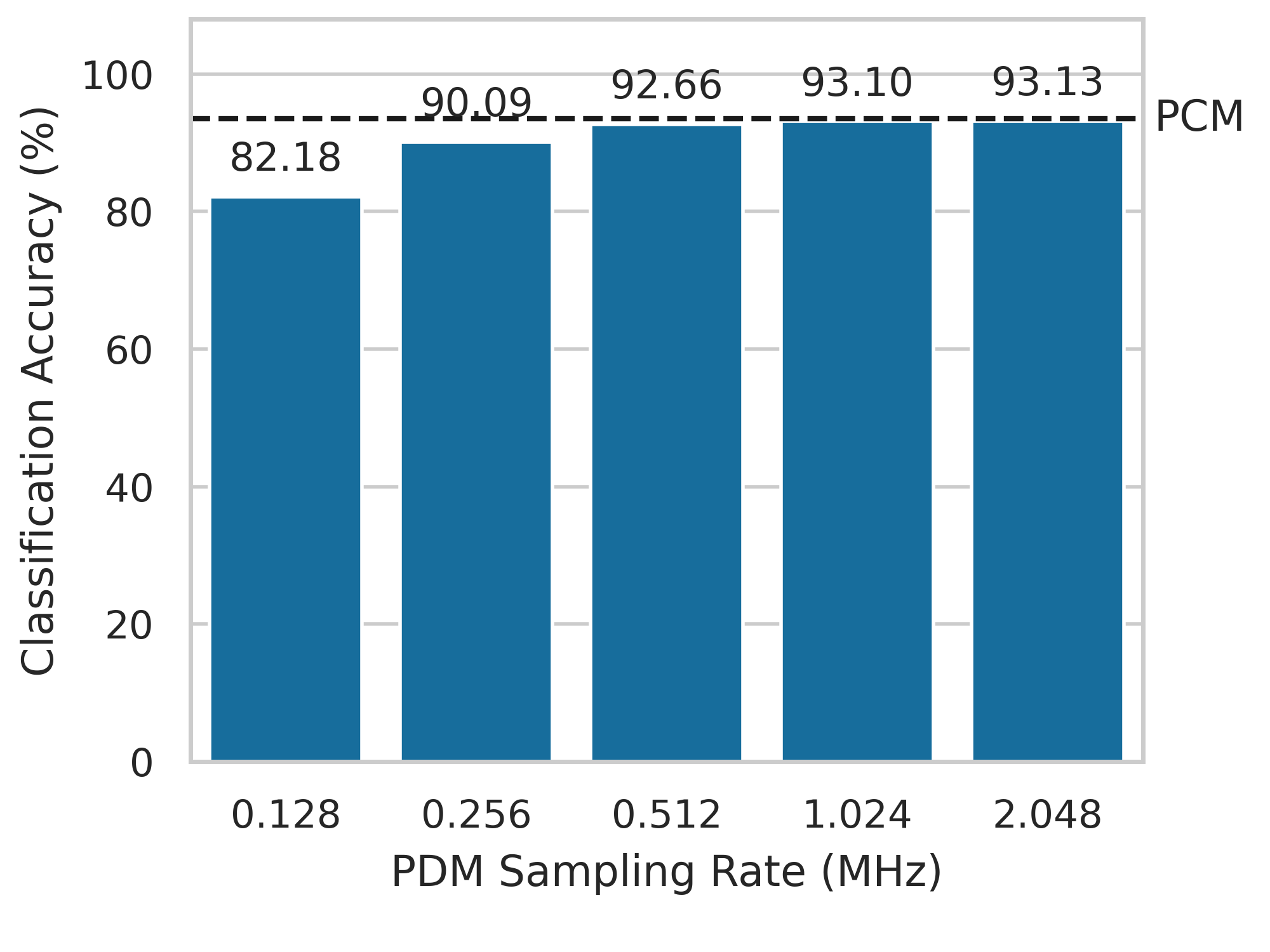}
    \caption{The performance of the LegT SSM with $\theta=$ \SI{2}{\ms} and $H=32$ on the Google Speech Commands test set encoded using a single-bit second-order $\Sigma\Delta$ modulator with OSRs varying from 8-128 (\SI{128}{\kHz}-\SI{2}{\MHz}). The horizontal dotted line indicates the test-time accuracy of 93.53\%, which was obtained on the PCM data.}
    \label{fig:ks-pdm-generalization}
\end{figure}

\subsubsection{Effect of PDM Sampling Rate on the Classification Accuracy}
Finally, we test the robustness of our proposed architecture to the various operating modes of a PDM microphone. Similar to the second experiment, we chose the lowest-latency and least computationally expensive SSM configuration. Thus, the state dimension $H$ and the window length $\theta$ are set to 32 and \SI{2}{\ms} respectively. The results are shown in Figure \ref{fig:ks-pdm-generalization}, which compares the test-time classification accuracy for each PDM sampling rate with the test-time PCM classification performance (black dotted line). The Figure shows that our method effectively generalizes to PDM data without ever having seen it during training. Indeed, we maintain over 92\% classification accuracy for PDM OSRs sampling rates above \SI{512}{\kHz}. This represents less than a 4\% drop in performance across these OSRs, highlighting that even the simplest SSM model can function across all low-power and standard/high-resolution PDM operating modes.

\subsection{Speech Enhancement Experiments}\label{sec:se-results}
We next conduct the same three experiments on a speech enhancement task. The reported metrics are the PESQ score \cite{rix_perceptual_2001} and the Short-time Objective Intelligibility (STOI) \cite{taal_stoi_2010}, which estimate the quality and intelligibility of the enhanced speech, respectively.
\subsubsection{Effect of Varying the Memory Window Length $\theta$ and State Dimension $H$ on Speech Enhancement Performance}
\begin{figure}[t!]
    \centering
    \includegraphics[width=\linewidth]{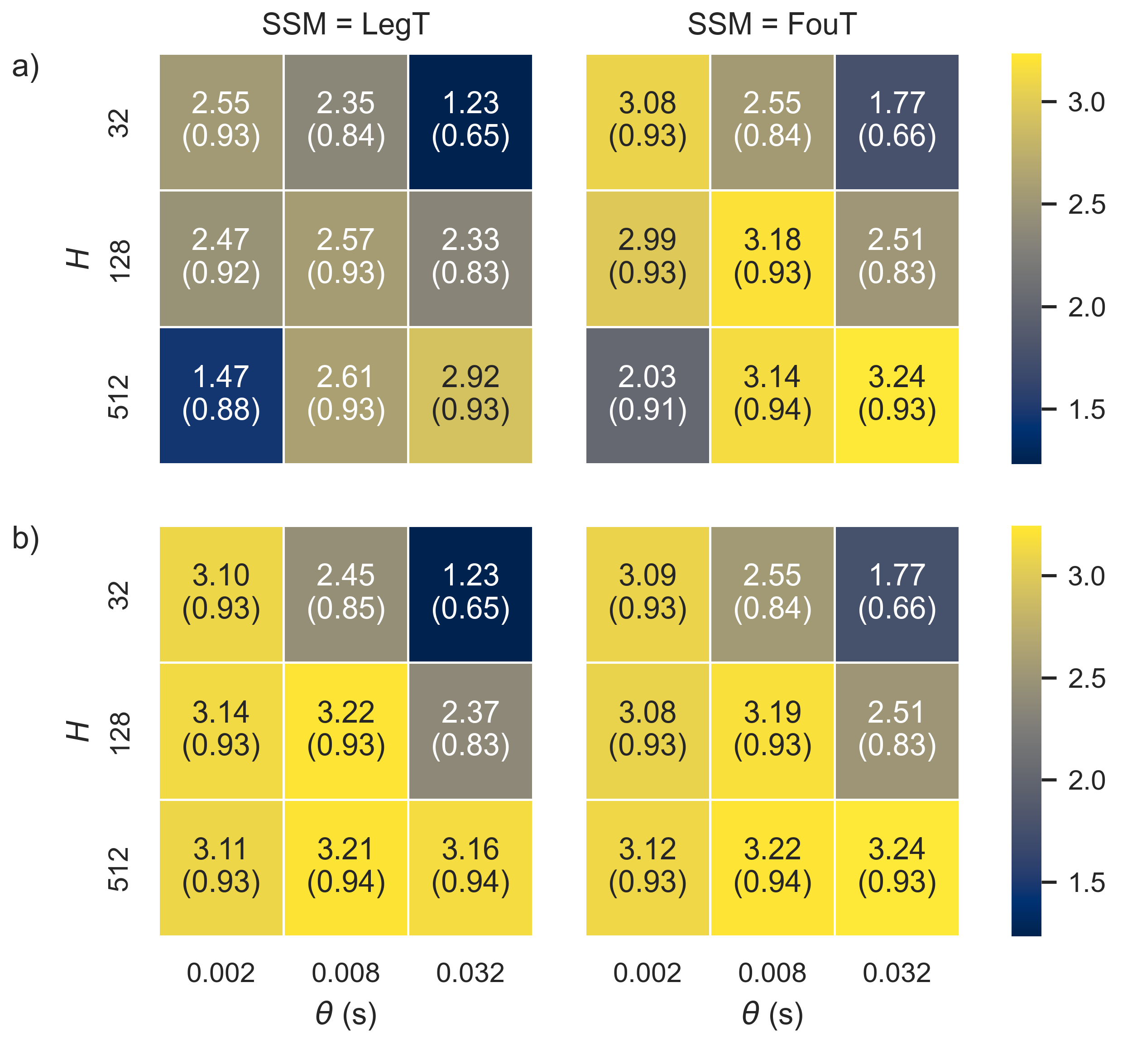}
    \caption{Effect of $\theta$ and $H$ on speech enhancement performance on the VoiceBank+DEMAND test set. For both SSM families, the decimation factor $K_{PDM}$ was fixed at 128, the window size $\theta$ was varied from \SI{2}{\ms} to \SI{32}{\ms}, and the state dimension H was varied from 32 to 512. The number at the top represents the PESQ score for a given combination, and the number at the bottom represents the STOI score. a) The results obtained on the test set encoded using a single-bit second-order $\Sigma\Delta$ modulator with an OSR of 128 (\SI{2}{\MHz}). b) The results obtained on test set encoded using \SI{16}{\kHz} PCM.}
    \label{fig:pdm-se-results-window}
\end{figure}

The results of varying $\theta$ and $H$ are presented in Figure \ref{fig:pdm-se-results-window}. From the Figure, we see that the best configuration is the FouT SSM with $\theta=$ \SI{32}{\ms} and $H=512$, which achieves a PESQ score of 3.24 and a STOI score of 93\%. Remarkably, even the configuration using $\theta=$ \SI{2}{\ms} and $H=32$ achieves a PESQ score of 3.08 and a STOI score of 93\%, which are very similar to the 3.07 PESQ score and 94\% STOI score obtained by the original LiSenNet architecture on PCM data \cite{yan_lisennet_2025}. These results show that our method enables natural generalization to PDM data without any loss of performance compared to the original network on PCM data.

Additionally, we note that, contrary to the keyword spotting task, the choices of FouT versus LegT and their state dimension $H$ appear to be significantly more important for the speech enhancement task. In fact, Figure \ref{fig:pdm-se-results-window}a) shows that the FouT SSM consistently outperforms the LegT SSM across all values of $H$ and $\theta$ and that the performance degrades significantly when $\bar{\theta}_{PCM}$ differs from $H$. This is not the case when the models are tested on PCM data as depicted in \ref{fig:pdm-se-results-window}b). In this case, the performance of the two SSMs is much more similar, and the LegT SSM even sometimes outperforms the FouT SSM.

\begin{figure}[b!]
    \centering
    \includegraphics[width=\linewidth]{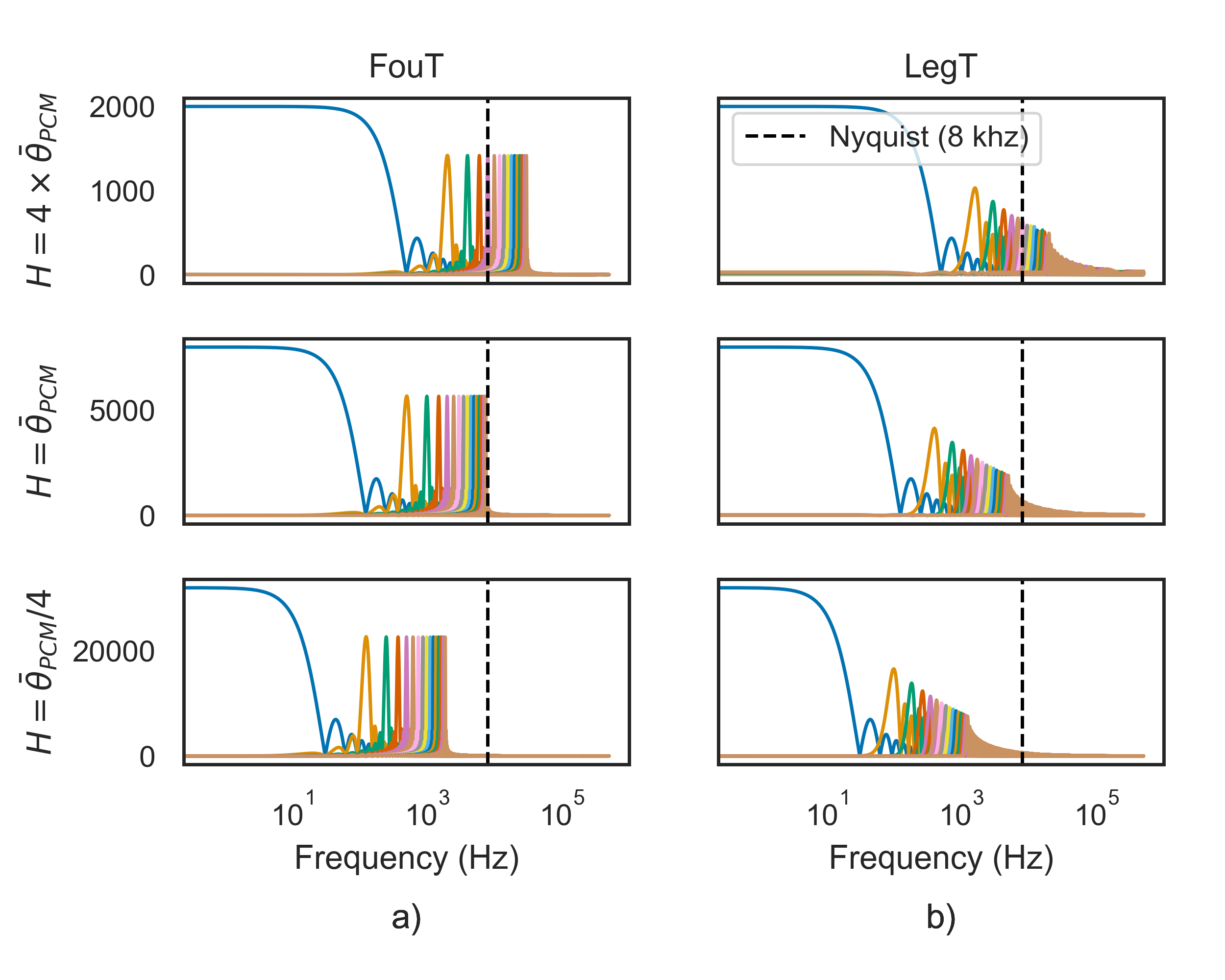}
    \caption{The frequency spectra of the basis vectors for both the FouT (left) and LegT (right) SSMs with $H=32$ and using window lengths of \SI{2}{\ms} (top), \SI{8}{\ms} (middle), and \SI{32}{\ms} (bottom). The black dotted line represents the PCM Nyquist frequency of \SI{8}{\kHz}.}
    \label{fig:basis-aliasing}
\end{figure}

These results can be explained by inspecting the passband of both SSMs for different values of $\theta$ depicted in Figure \ref{fig:basis-aliasing}. On the one hand, when the $\bar{\theta}_{PCM}$ is smaller than $H$, the frequency spectrum of the basis vectors actually spans frequencies higher than the PCM Nyquist frequency. Thus, when encoding PDM data, these bases will respond to quantization noise above Nyquist that was never seen during training. On the other hand, when $\bar{\theta}_{PCM}$ is larger than $H$, the basis vectors are not able to span the whole frequency range from 0-Nyquist. This creates an inherent lowpass effect in the encoding SSM and prevents the downstream modules from encoding the higher frequency content of the input speech signals.

When $H$ is equal to $\bar{\theta}_{PCM}$ (i.e., the diagonal in Figure \ref{fig:pdm-se-results-window}), the FouT basis vectors span the whole frequency range from 0-Nyquist without going over it. However, the LegT basis vectors still span frequencies higher than the PCM Nyquist frequency. This difference arises because the FouT basis vectors' frequency content is narrowband, and the number of encoded frequencies depends solely on the state dimension $H$. Therefore, the basis vectors encode the same frequencies whether the input is PCM or PDM. In contrast, the Legendre polynomials' frequency content has a much wider bandwidth. This doesn't pose a problem when testing on PCM data, as their frequency content is limited to the Nyquist range. However, when testing on PDM data, the frequency content now spans the range 0 to OSR$\times$Nyquist, with high-energy quantization noise occupying the spectrum above the PCM Nyquist. Therefore, the neural network is suddenly exposed to new frequencies it has never seen during training, and that contain only quantization noise. Moreover, when we reconstruct the PCM speech by projecting the PDM coefficients on the Legendre polynomials evaluated at the PCM sampling rate during the synthesis stage, these frequencies are aliased into the signal band, resulting in much poorer reconstruction than the FouT SSM can achieve.

\begin{figure}[t!]
    \centering
    \includegraphics[width=\linewidth]{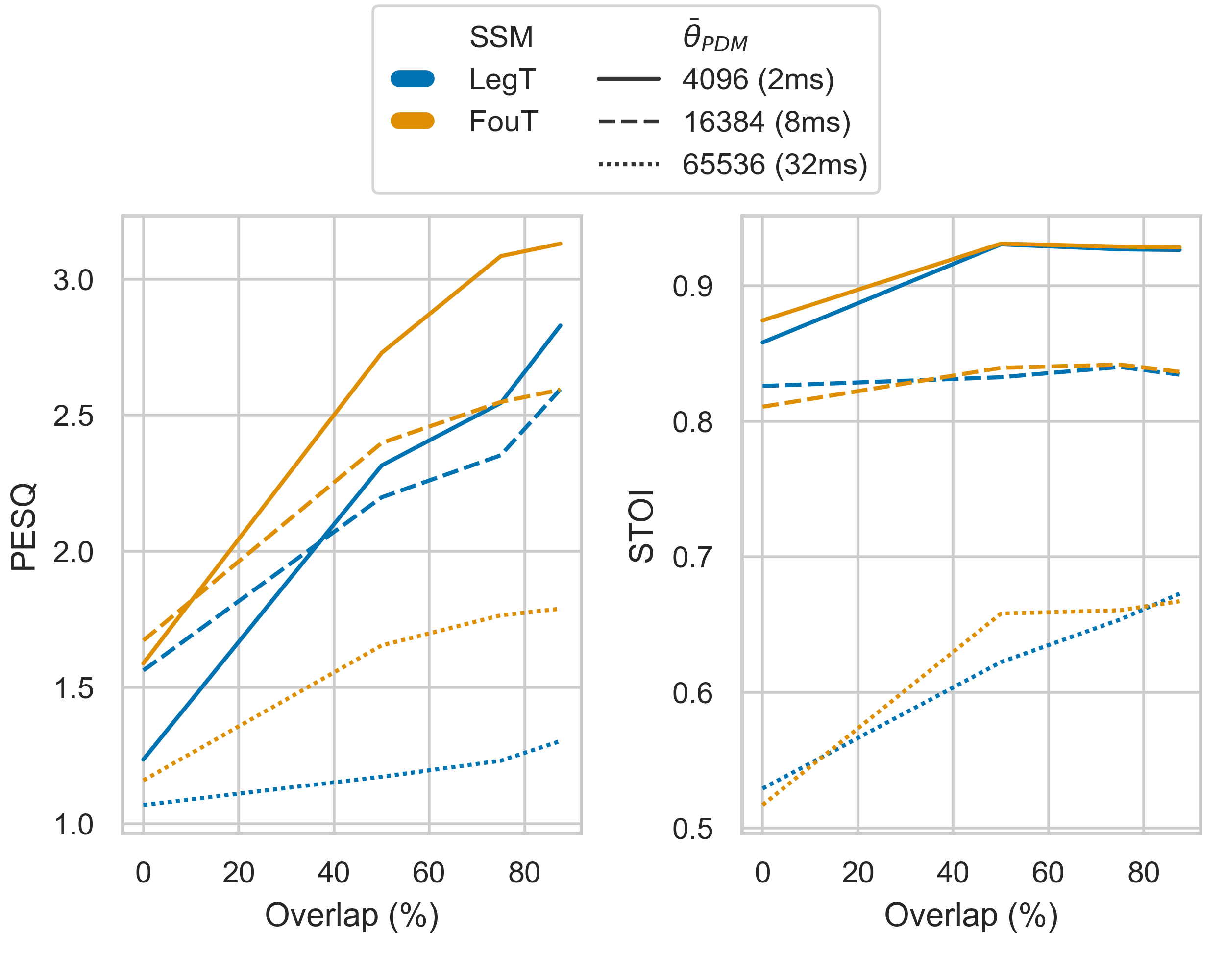}
    \caption{Effect of the overlap and the memory window length $\bar{\theta}_{PDM}$ on speech enhancement performance on the VoiceBank+DEMAND test set encoded using a single-bit second-order $\Sigma\Delta$ modulator with an OSR of 128 (\SI{2}{\MHz}). For both SSM families, the overlap was varied from 0-87.5\% and $\bar{\theta}_{PDM}$ was varied from 4,096 to 65,536.}
    \label{fig:pdm-se-results-decimation}
\end{figure}

\subsubsection{Effect of the Decimation Factor $K_{PDM}$ and the Memory Window Length $\bar{\theta}_{PDM}$ on Speech Enhancement Performance}\label{sec:se-downsampling}

We next move to testing the effect of the decimation factor $K_{PDM}$ on the speech enhancement performance. In this experiment, we vary the decimation factor as a fraction of the memory window length akin to the overlap percentage in standard OLA processing. This allows us to study the impact of the overlap percentage between windows on the reconstruction accuracy. We also fix the state dimension $H$ to 32 as this represents the configuration with the lowest computational complexity, thus best suited for low-power, always-on devices. The results are presented in Figure \ref{fig:pdm-se-results-decimation}. The figure shows that performance correlates with the overlap between consecutive windows. In fact, greater overlap enables the network to denoise speech more efficiently. Remarkably, the FouT SSM achieves a PESQ score of 3.13 and 3.08 for overlaps of 87.5\% and 75\% respectively. These scores are similar or greater than the 3.07 PESQ score obtained by the original LiSenNet network. This is ideal for low-latency systems such as hearing aids which require delays under \SI{6}{\ms} \cite{Stone2008TolerableHearingAidDelaysV}. In fact, if we define the system's algorithmic latency as the time it takes to fully reconstruct an output speech sample $y[n]$, then a 75\% overlap implies that three output frames are required to fully reconstruct $y[n]$ and that an output frame is produced at every \SI{1.5}{\ms}. Therefore, the resulting in a total algorithmic latency is \SI{4.5}{\ms}.

\begin{figure}[t!]
    \centering
    \includegraphics[width=\linewidth]{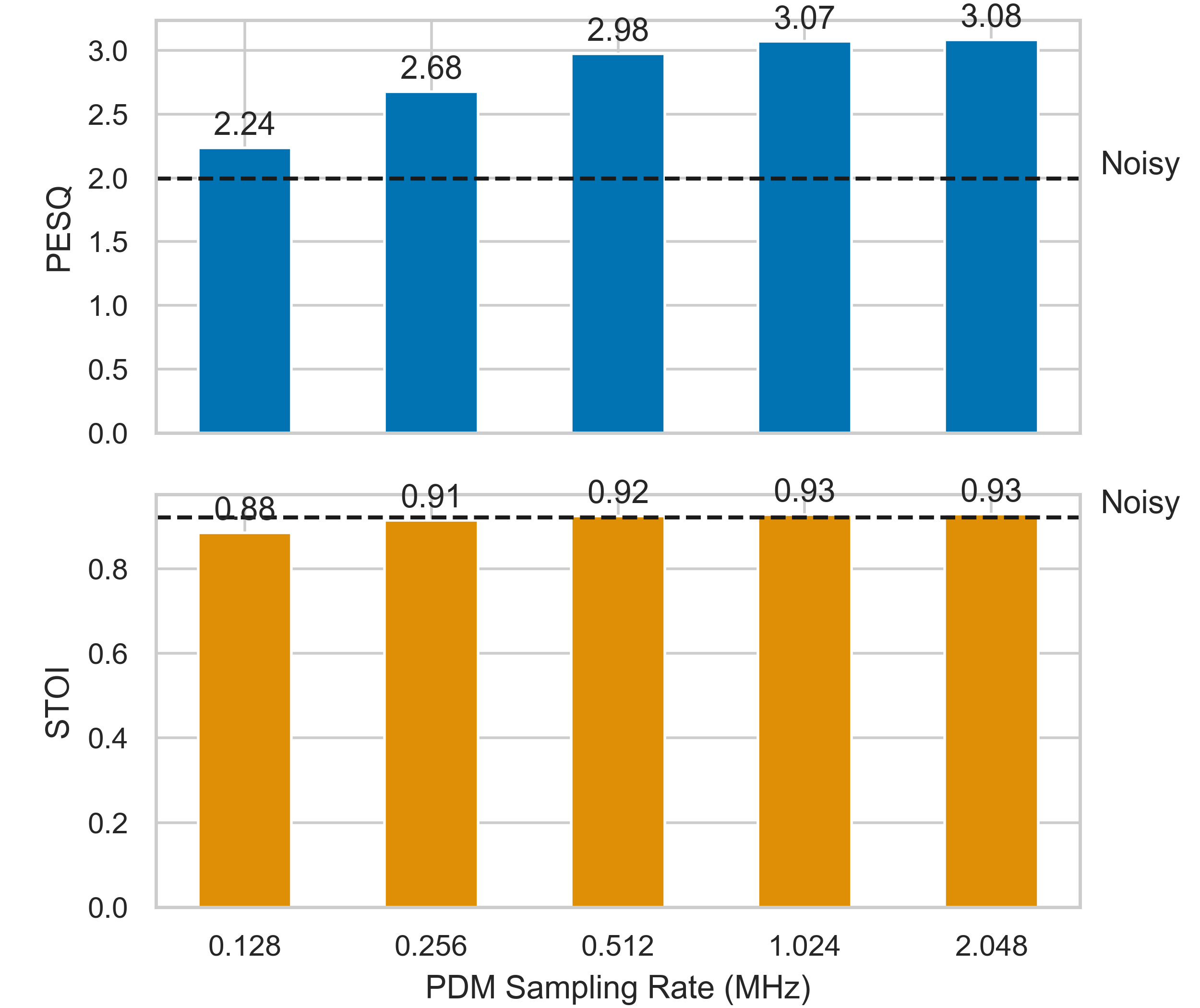}
    \caption{The performance of the FouT SSM with a \SI{2}{\ms} window size, a state dimension of 32, and an overlap of 75\% on the VoiceBank+DEMAND test set encoded using a single-bit second-order $\Sigma\Delta$ modulator with a OSRs varying from 8-128 (\SI{128}{\kHz}-\SI{2}{\MHz}). The black dotted lines represent the PESQ/STOI scores of 2.0 and 0.92 obtained by comparing the noisy speech versus the clean speech on the same test set.}
    \label{fig:pdm-se-results-config}
\end{figure}

\subsubsection{Effect of PDM Sampling Rate on Speech Enhancement Performance}
Finally, we again test the FouT SSM with a \SI{2}{\ms} memory window and a 32-dimensional state across all PDM sampling rates. Figure \ref{fig:pdm-se-results-config} presents the results. The noisy speech PESQ score is 2.0, while our architecture's lowest PESQ score is 2.24, at the lowest PDM sampling rate of \SI{128}{\kHz}. Similarly, the noisy speech STOI score is 0.92, which our model already matches at \SI{512}{\kHz}. These results are promising, since the PDM low-power mode typically operates between \SI{350}{\kHz} and \SI{1}{\MHz} and the standard/high-resolution mode typically operates over \SI{1}{\MHz} \cite{knowles_sph0641lm4h1}: our architecture thus offers speech enhancement benefits in both low-power and standard/high-resolution operating modes, without any finetuning. In contrast, previous approaches that used test-time PDM OSRs different from training-time PDM OSRs saw a decrease of almost 10\% for only a 50\% decrease in PDM OSR \cite{arnaud_yarga_end--end_2025}.

\section{Conclusion}\label{sec:conclusion}
This work presented an efficient PDM speech-processing architecture that can be trained solely on PCM sequences and can generalize across low-power and standard/high-resolution PDM operating modes without any finetuning. We achieved this by using a SSM encoder layer that produces a modulation- and sampling-rate-invariant representation of the input speech. Prior PDM approaches typically required training directly on long PDM sequences and re-training for each target OSR to avoid significant performance loss \cite{vitolo_new_2023, arnaud_yarga_end--end_2025, yarga24_interspeech}. In contrast, we have shown that our architecture can easily adapt SOTA keyword spotting and speech enhancement DNNs such that they can be trained on short \SI{16}{\kHz} PCM sequences and deployed directly to PDM data at various sampling rates without retraining. In the low-power operating range (\SI{512}{\kHz}), our most computationally efficient architecture maintains 92.66\% keyword-spotting accuracy and a PESQ/STOI of 2.98/0.92. In the standard operating range (\SI{2}{\MHz}), it maintains 93.13\% accuracy and PESQ/STOI of 3.08/0.93, which are all within 0.5\% of the PCM baseline, thus confirming that performance is robust to the modulation strategy across both tested regimes.

Nevertheless, our speech enhancement experiments raised a critical question: is it possible to increase the number of basis vectors in the SSM such that there are more basis vectors than sample points in the memory window without encoding frequencies higher than the Nyquist frequency? This would be analoguous to standard STFT processing where the STFT can be padded to increase frequency resolution. Therefore, future work will investigate how the FouT basis vectors could be focused to represent the frequencies under the PCM Nyquist frequency, analogous to the STFT, thereby increasing the frequency resolution of the FouT SSM under the PCM Nyquist frequency and possibly leading to better speech enhancement performance.

Moreover, one of the main motivations for this work was to design an algorithm that could adapt to changes in the PDM sampling-rates in real-time. This would allow always-on devices such as hearing-aids to maximize their battery life by using the low-power setting of the PDM microphone when there is no active speech detected, and changing to the standard mode in the opposite case. To this end, we have demonstrated that the proposed architecture handles various PDM sampling-rates effectively, establishing a strong foundation for real-time adaptability. Building on this, future work will implement the algorithm on an embedded platform such as a field-programmable gate array (FPGA) connected to a PDM microphone, enabling us to validate the proposed architecture's adaptability to sudden changes in the PDM sample rates.

\bibliographystyle{IEEEtran}
\bibliography{bibliography}
\end{document}